\documentclass[12pt]{article}
\pdfoutput=1

\usepackage{amsmath,amssymb,amscd}
\usepackage{listings}
\usepackage{caption}
\usepackage{dsfont}
\usepackage{slashed}
\usepackage{color}

\usepackage[pdftex]{graphicx}
\usepackage{epstopdf}
\usepackage{subfigure}
\usepackage{epsfig}
\usepackage{listings}
\usepackage{caption}
\usepackage{cite}

\usepackage{multirow}

\newcommand{\bea}{\begin{align}}
\newcommand{\eea}{\end{align}}
\newcommand{\beq}{\begin{equation}}
\newcommand{\eeq}{\end{equation}}
\newcommand{\nbea}{\begin{align*}}
\newcommand{\neea}{\end{align*}}
\newcommand{\nbeq}{\begin{equation*}}
\newcommand{\neeq}{\end{equation*}}
\newcommand{\bear}{\begin{eqnarray}}  
\newcommand{\eear}{\end{eqnarray}}  

 \def\lra#1{\overset{\text{\scriptsize$\leftrightarrow$}}{#1}}

\numberwithin{equation}{section}
\begin{document}

\begin{titlepage}

\pagestyle{empty}

\baselineskip=21pt
{\small
\rightline{KCL-PH-TH/2017-04, CERN-PH-TH/2017-009}
\rightline{Cavendish-HEP-17/01, DAMTP-2017-01}
}
\vskip 0.4in

\begin{center}

{\Large {\bf Dimension-6 Operator Analysis of the CLIC Sensitivity to New Physics}}

\vskip 0.4in

 {\bf John~Ellis~}$^{1,2}$.
 {\bf Philipp~Roloff~}$^3$,
 {\bf Ver\'onica~Sanz~}$^4$ 
and {\bf Tevong~You~}$^5$

\vskip 0.4in

{\small {\it
$^1${Theoretical Particle Physics and Cosmology Group, Physics Department, \\
King's College London, London WC2R 2LS, UK}\\
\vspace{0.25cm}
$^2${Theoretical Physics Department, CERN, CH-1211 Geneva 23, Switzerland} \\
\vspace{0.25cm}
$^3${Experimental Physics Department, CERN, CH-1211 Geneva 23, Switzerland} \\
\vspace{0.25cm}
$^4${Department of Physics and Astronomy, University of Sussex, Brighton BN1 9QH, UK} \\
\vspace{0.25cm}
$^5${Cavendish Laboratory, University of Cambridge, J.J. Thomson Avenue, \\
Cambridge, CB3 0HE, UK; \\
\vspace{-0.25cm}
DAMTP, University of Cambridge, Wilberforce Road, Cambridge, CB3 0WA, UK}
}}

\vskip 0.4in

{\bf Abstract}

\end{center}

\baselineskip=18pt \noindent

%%%%%%%%%%%%%%%%%%%%%%%%%%%%%%%%%%%%%%%%%%%%%%%%%

{%\small

We estimate the possible accuracies of measurements at the proposed CLIC $e^+e^-$ collider 
of Higgs and $W^+W^-$ production at centre-of-mass energies up to 
3~TeV, incorporating also Higgsstrahlung projections at higher energies that had not been considered
previously, and use them to explore the prospective CLIC sensitivities to decoupled new physics. 
We present the resulting constraints on the Wilson coefficients of 
dimension-6 operators in a model-independent approach based on the Standard Model 
effective field theory (SM EFT). The higher centre-of-mass energy of CLIC,
compared to other projects such as the ILC and CEPC, gives it greater
sensitivity to the coefficients of some of the operators we study. We find that
CLIC Higgs measurements may be sensitive to new physics scales $\Lambda = \mathcal{O}(10)$~TeV 
for individual operators, reduced to $\mathcal{O}(1)$ TeV sensitivity for a global fit marginalising 
over the coefficients of all contributing operators. 
We give some examples of the corresponding
prospective constraints on specific scenarios for physics beyond the SM, including stop quarks and the dilaton/radion.
}

%%%%%%%%%%%%%%%%%%%%%%%%%%%%%%%%%%%%%%%%%%%%%%%%

\vskip 0.4in

\leftline{January 2017}

\end{titlepage}

\newpage

%\tableofcontents

%*******************************************************************************************
\section{Introduction}

In view of the fact that, so far, the couplings of the known particles are consistent with those
predicted in the Standard Model (SM), and in the absence of any evidence at the LHC or
elsewhere of any particles beyond those in the SM, it is natural to assume that any new 
physics must involve massive particles that are decoupled at the energies explored so 
far~\cite{decoupling}. Such models of new physics may be explored using the SM Effective
Field Theory (SM EFT), which provides a model-independent parametrisation of the low-energy
effects of such new physics via higher-dimensional operators constructed out of SM fields~\cite{buchmullerwyler, GIMR}.
In collider physics, operators of dimension 6 are typically the most important, since the dimension-5
Weinberg neutrino-mass operator~\cite{weinbergoperator} does not play a r\^ole~\footnote{For
instances where dimension-8 operators may dominate, see for example Refs.~\cite{dim8, EFTvalidity}.}. Such operators
may be generated at either tree or loop level, and in the latter case the matching to ultraviolet (UV) 
models is easily achieved by the universal one-loop effective action~\cite{universal}.

Data from the LHC, LEP and SLC have already been analysed using the SM EFT~\footnote{See Ref.~\cite{review} and references therein for a review of the SM EFT.}~\cite{earlyeft,
HanSkiba, eftconstraints, ciuchinietal, PomarolRiva, ESY, falkowskiriva, berthiertrott, 
flavourfulEFT, EFThiggsreview, higgslegacyEFT, topquarkEFT, TGCEFT, pushinghiggsEFT} in various choices of operator 
bases~\footnote{The 
{\tt Rosetta} tool~\cite{rosetta} may be used to translate between different operator bases
used in the literature.}. In particular, three of us (JE, VS and TY) 
have published a global analysis of dimension-6 operators in the SM EFT~\cite{ESY}, 
providing 95\% CL ranges for their coefficients when
marginalising over the possible coefficients of all contributing operators and when the operators
are switched on individually, and similar global fits to dimension-6 operators can be found in, 
for example, Refs.~\cite{eftconstraints, PomarolRiva, higgslegacyEFT}.
The prospective sensitivities of future accelerators such as the HL-LHC, the ILC and
FCC-ee have also been estimated~\cite{ILCwhitepaper, ILCquantum, ILC, ILCop, TLEPdesigndoc, futurestudies,EY}.

We show in this paper that the Compact Linear Collider (CLIC) proposal for
an $e^+ e^-$ collider with a centre-of-mass energy $\le 3$ TeV~\cite{CLIC} has a unique advantage for probing the
coefficients of dimension-6 operators in the SM EFT, and hence constraining scenarios for
possible new physics beyond the SM. This is because the relative contributions of some dimension-6 operators
to scattering amplitudes grow rapidly with the energy $E$, since interferences with SM amplitudes grow
like $E^2$ relative to the latter, conferring a competitive advantage on a higher-energy collider that can attain
measurement accuracies comparable with a lower-energy $e^+ e^-$ collider such as the ILC,
FCC-ee or CEPC. 

Previous studies of CLIC projections for Higgs physics focused on the $WW$ fusion production mode that
dominates Higgs production at higher energies~\cite{CLICH}. Here we point out that associated $HZ$ production 
at 1.4 and 3 TeV could be important for indirect new physics searches, despite the lower statistics. 
We had previously shown that the analogous Higgsstrahlung production mode at the LHC is
particularly sensitive to certain dimension-6 operators whose contributions to the cross-section
and kinematic distributions grow with energy~\cite{ESY} (See also Ref.~\cite{compositefermions})~\footnote{The energy growth of dimension-6 operators has also been used to place constraints from Drell-Yan processes at hadron colliders~\cite{drellyan}.}. For the same reason, inclusion of
associated $HZ$ production has a dramatic effect on the SM EFT fit at CLIC. 

The layout of this paper is as follows.
 
In Section~2 of this paper we recall relevant aspects of the SM EFT, highlighting the operators
that contribute to processes measurable at CLIC at high energies, such as Higgs production
via both $e^+ e^- \to HZ$ associated production and vector-boson fusion (VBF), and
$e^+ e^- \to W^+ W^-$, which constrains SM EFT coefficients via triple-gauge couplings (TGCs). 
We also recall that there are some combinations of coefficients of
dimension-6 operators that are particularly strongly constrained by electroweak precision tests (EWPTs).

In Section~3 of this paper we provide indicative estimates of the accuracy with which the
cross sections for $e^+ e^- \to HZ$ and $e^+ e^- \to W^+ W^-$
might be measurable at CLIC at centre-of-mass energies ranging from 350~GeV through 1.4~TeV
to 3~TeV. We emphasise that these estimates are not yet based on detailed
simulations, which are now being prepared by the CLICdp Collaboration.
We also repurpose previous ILC studies to estimate how accurately Higgs
observables could be measured at CLIC at 350~GeV.

In Section~4 we use these estimates to provide projections of the possible CLIC sensitivity
to the coefficients of dimension-6 operators, treated both individually and marginalised in global fits.
We provide numerical support for the claim made above that the higher-energy CLIC
measurements could provide significantly improved sensitivity to some of these coefficients,
precisely by virtue of the growth in their contributions to cross sections relative to calculations
in the SM. We find that the dependence on certain operator coefficients may grow by a factor 
$\mathcal{O}(100)$ between 350 GeV and 3 TeV, leading to individual 95\% CL sensitivities to new physics that
may reach up $\sim 10$ TeV via certain operators whose contributions to cross-sections grow with energy.

Section~5 explores the applicability and utility of these CLIC estimates in the contexts of some
specific scenarios for physics beyond the SM, namely stop squarks in the minimal supersymmetric
extension of the SM (MSSM) and dilaton/radion models.

Finally, Section~6 summarises our conclusions, emphasising the importance of detailed
simulations of CLIC data at high energies to justify and refine the estimates we make of
the accuracies with which the $e^+ e^- \to HZ$, $e^+ e^- \to H \nu {\bar \nu}$ and 
$e^+ e^- \to W^+ W^-$ cross sections should be measurable at CLIC.

\section{The Standard Model Effective Field Theory}

In the Standard Model Effective Field Theory (SM EFT) it is assumed that all the known
particles have exactly the same renormalisable couplings as predicted in the SM, but that these
are supplemented by interactions characterised by higher-dimensional operators constructed out of SM fields. Thus,
one considers all possible combinations of SM fields of a given dimensionality that are consistent with the SM
SU(3)$_c \times$SU(2)$_L \times$U(1)$_Y$ gauge symmetries and Lorentz invariance.
We focus here on the leading lepton-number-conserving dimension $d \ge 6$ operators,
whose Wilson coefficients could be generated by decoupled new physics beyond the SM:
\begin{equation}
\mathcal{L}_\text{SMEFT} = \mathcal{L}_\text{SM} + \sum_i \frac{c_i}{\Lambda^2}\mathcal{O}_i 	\, .
\label{SMEFT}
\end{equation}
The effects of operators with dimensions $d > 6$ are sub-dominant in such a decouplings
scenario -- with some exceptions~\cite{dim8} -- justifying our focus on the dimension-6 terms in the SM EFT Lagrangian:
For the purposes of our analysis, we express the dimension-6 operators $\mathcal{O}_i$ in the 
basis of Ref.~\cite{PomarolRiva}. The dimensionful parameter $\Lambda$ in (\ref{SMEFT}) reflects the scale of 
the new decoupled physics, and the coefficients $c_i$ are model-dependent. 
We assume CP conservation and a flavour-blind structure for the operators involving SM fermions,
so that the operators relevant for the precision electroweak, Higgs and $e^+ e^- \to W^+ W^-$
observables that we include in our fits are those listed in Table \ref{tab:dim6}.

\vspace{0.5cm}
\begin{table}[h]
{\small
\begin{center}
\begin{tabular}
{ | c | c | c | } 
\hline
EWPTs & Higgs Physics & TGCs \\
\hline
\multicolumn{3}{| c |}{ ${\mathcal O}_W=\frac{ig}{2}\left( H^\dagger  \sigma^a \lra {D^\mu} H \right )D^\nu  W_{\mu \nu}^a$  } \\
\hline
\multicolumn{2}{| c |}{ ${\mathcal O}_B=\frac{ig'}{2}\left( H^\dagger  \lra {D^\mu} H \right )\partial^\nu  B_{\mu \nu}$  } & ${\mathcal O}_{3W}= g \frac{\epsilon_{abc}}{3!} W^{a\, \nu}_{\mu}W^{b}_{\nu\rho}W^{c\, \rho\mu}$	\\
\hline
${\cal O}_T=\frac{1}{2}\left (H^\dagger {\lra{D}_\mu} H\right)^2$ & \multicolumn{2}{| c |}{ ${\mathcal O}_{HW}=i g(D^\mu H)^\dagger\sigma^a(D^\nu H)W^a_{\mu\nu}$ } \\
\hline
$\mathcal{O}_{LL}^{(3)\, l}=( \bar L_L \sigma^a\gamma^\mu L_L)\, (\bar L_L \sigma^a\gamma_\mu L_L)$ & \multicolumn{2}{| c |}{$ {\mathcal O}_{HB}=i g^\prime(D^\mu H)^\dagger(D^\nu H)B_{\mu\nu}$ } \\
\hline
${\mathcal O}_R^e = (i H^\dagger {\lra { D_\mu}} H)( \bar e_R\gamma^\mu e_R)$ & ${\mathcal O}_{g}=g_s^2 |H|^2 G_{\mu\nu}^A G^{A\mu\nu}$ &  \\
\hline
${\cal O}_{R}^u = (i H^\dagger {\lra { D_\mu}} H)( \bar u_R\gamma^\mu u_R)$ & ${\mathcal O}_{\gamma}={g}^{\prime 2} |H|^2 B_{\mu\nu}B^{\mu\nu}$ & \\
\hline
${\cal O}_{R}^d = (i H^\dagger {\lra { D_\mu}} H)( \bar d_R\gamma^\mu d_R)$ & ${\mathcal O}_H=\frac{1}{2}(\partial^\mu |H|^2)^2$ & \\
\hline
${\cal O}_{L}^{(3)\, q}=(i H^\dagger \sigma^a {\lra { D_\mu}} H)( \bar Q_L\sigma^a\gamma^\mu Q_L)$  & ${\mathcal O}_{f}   =y_f |H|^2    \bar{F}_L H^{(c)} f_R + \text{h.c.}$ & \\
\hline
${\cal O}_{L}^q=(i H^\dagger {\lra { D_\mu}} H)( \bar Q_L\gamma^\mu Q_L)$  & $\mathcal{O}_6 = \lambda|H|^6$ & \\ 
\hline
\end{tabular}
\end{center}
}
\caption{\it List of pertinent CP-even dimension-6 operators in the basis~\cite{PomarolRiva} that we use. In each case
we recall the categories of observables that provide the greatest sensitivities to the operator,
possibly in combinations with other operators.}
\label{tab:dim6}
\end{table}

As noted in Table~\ref{tab:dim6}, electroweak precision tests (EWPTs), 
notably those in the leptonic sector of $Z$-pole observables, 
provide the greatest sensitivities to the following (combinations of) dimension-6 operators:
\begin{equation}
\mathcal{L}_\text{dim-6}^\text{EWPT} \supset \frac{1}{2}\frac{(\bar{c}_W + \bar{c}_B)}{m_W^2} (\mathcal{O}_W + \mathcal{O}_B) + \frac{\bar{c}_T}{v^2}\mathcal{O}_T + \frac{\bar{c}^{(3)l}_{LL}}{v^2}\mathcal{O}^{(3)l}_{LL} + \frac{\bar{c}^e_R}{v^2}\mathcal{O}^e_R 	\, .
\label{EWPTs}
\end{equation}
In writing (\ref{EWPTs}), we have introduced coefficients $\bar{c}_i$ that differ from those in (\ref{SMEFT})
by ratios of the squares of an electroweak scale $M$ to the effective scale $\Lambda$ of new physics:
\begin{equation}
\bar{c}_i = c_i \frac{M^2}{\Lambda^2} \, ,
\label{cbarc}
\end{equation}
where $M \equiv m_W$ for the combination $\mathcal{O}_W + \mathcal{O}_B$, and
$M \equiv v$ for the other operators. 

As also noted in Table~\ref{tab:dim6}, the dimension-6 operators (and their linear combinations) relevant to the Higgs and 
triple-gauge coupling (TGC) measurements used in our fits are
\begin{align}
\mathcal{L}_\text{dim-6}^\text{Higgs+TGC} &\supset \frac{1}{2}\frac{(\bar{c}_W - \bar{c}_B)}{m_W^2}(\mathcal{O}_W - \mathcal{O}_B) +\frac{\bar{c}_{HW}}{m_W^2}\mathcal{O}_{HW} + \frac{\bar{c}_{HB}}{m_W^2}\mathcal{O}_{HB} + \frac{\bar{c}_g}{m_W^2}\mathcal{O}_{g} + \frac{\bar{c}_\gamma}{m_W^2}\mathcal{O}_\gamma  \nonumber \\
&\quad \quad + \frac{\bar{c}_{3W}}{m_W^2}\mathcal{O}_{3W} + \frac{\bar{c}_H}{v^2}\mathcal{O}_H + \frac{\bar{c}_f}{v^2}\mathcal{O}_f 	\, .
\label{HTGCs}
\end{align}
Since the EWPTs can constrain very strongly the linear combination $\bar{c}_W + \bar{c}_B$,
we assume $\bar{c}_B = -\bar{c}_W$ when fitting Higgs and TGC measurements. 
When marginalising over the effects of all operators in our global fits this assumption holds
less well for the TGCs~\cite{TCGs-caveat}, but is sufficient for our projections where we are mainly interested in
a first estimate of the CLIC sensitivity to the scale of new physics~\footnote{We note that the constraints
on the individual coefficients of single operators switched on one at a time typically show a high level of 
sensitivity to new physics, but one typically expects several operators to be generated when integrating out heavy particles
in any specific scenario for new physics~\cite{integrating-out}.}.

Our fit constrains the coefficients at the respective centre-of-mass energy scales $E$ at which they are measured:
$c_i \equiv c_i(E)$, which
are related to their values at the matching scale, $c_i(\Lambda)$, by renormalisation-group
equations (RGEs) that we do not consider here~\cite{RGE}. Also, we neglect dimension-8 
and higher-order operators in our analysis,
as well as the four-fermion operators that do not interfere with the SM amplitudes~\cite{HanSkiba}~\footnote{On 
the other hand, we do include $\bar{c}^{(3)l}_{LL}$ when analysing the EWPTs, 
which modifies the input parameter $G_F$.}, whose effects on $Z$-pole measurements are
of the same order in $\Lambda$ (or $M$) as dimension-8 operators. It was pointed out in~\cite{berthiertrott}
that these operators and theory uncertainties that we omit could be relevant for $\Lambda \lesssim 3$ TeV. 
These effects and a consistent treatment of the SM EFT at one-loop level, including matching at 
one-loop~\cite{universal}, will become relevant for realistic fits 
as future precision data become available.

\section{CLIC Measurements}

The CLIC accelerator is foreseen to be built and operated in a staged approach with several 
centre-of-mass energy stages ranging from a few hundred GeV up to 3 TeV~\cite{CLIC}. The 
CLIC physics potential for the measurement of a wide range of Higgs boson properties has been investigated 
in detail based on full detector simulations~\cite{CLICH}. For our studies here, three energy stages 
at 350~GeV, 1.4~TeV and 3~TeV have been assumed.

\vspace{0.5cm}
\begin{table}[h]
{\small
\begin{center}
\begin{tabular}
{ | c | c | c | c | c | c | } 
\hline
\multicolumn{2}{|c|}{ CLIC at 350~GeV} & \multicolumn{2}{|c|}{ CLIC at 1.4~TeV} & \multicolumn{2}{|c|}{ CLIC at 3~TeV} \\
\hline
$\sigma_{H\nu_e\bar{\nu}_e}\text{BR}_{H\to b\bar{b}} $ & 1.9\% & $\sigma_{H\nu_e\bar{\nu}_e}\text{BR}_{H\to b\bar{b}} $ & 0.4\% & $\sigma_{H\nu_e\bar{\nu}_e}\text{BR}_{H\to b\bar{b}} $ & 0.3\% \\
\hline
$\sigma_{HZ}\text{BR}_{H\to c\bar{c}} $ & 10.3\% & $\sigma_{H\nu_e\bar{\nu}_e}\text{BR}_{H\to c\bar{c}} $ & 6.1\% & $\sigma_{H\nu_e\bar{\nu}_e}\text{BR}_{H\to c\bar{c}} $ & 6.9\%  \\
\hline
$\sigma_{HZ}\text{BR}_{H\to gg} $ & 4.5\% & $\sigma_{H\nu_e\bar{\nu}_e}\text{BR}_{H\to gg} $ & 5.0\% & $\sigma_{H\nu_e\bar{\nu}_e}\text{BR}_{H\to gg} $  & 4.3\%  \\
\hline
$\sigma_{HZ}\text{BR}_{H\to W^+W^-} $ & 5.1\% & $\sigma_{H\nu_e\bar{\nu}_e}\text{BR}_{H\to W^+W^-} $ & 1.0\% & $\sigma_{H\nu_e\bar{\nu}_e}\text{BR}_{H\to W^+W^-} $  & 0.7\%  \\
\hline
$\sigma_{HZ}\text{BR}_{H\to \tau\bar{\tau}} $ & 6.2\% & $\sigma_{H\nu_e\bar{\nu}_e}\text{BR}_{H\to \tau\bar{\tau}} $ & 4.3\% &  $\sigma_{H\nu_e\bar{\nu}_e}\text{BR}_{H\to \tau\bar{\tau}} $  & 4.4\% \\ 
\hline
$\sigma_{HZ}\text{BR}_{H\to b\bar{b}} $ & 0.84\% & $\sigma_{H\nu_e\bar{\nu}_e}\text{BR}_{H\to \gamma\gamma} $ & 15.0\% & $\sigma_{H\nu_e\bar{\nu}_e}\text{BR}_{H\to \gamma\gamma} $ & 10.0\% \\
\hline
 & & $\sigma_{H\nu_e\bar{\nu}_e}\text{BR}_{H\to Z\gamma} $ & 42.0\% & $\sigma_{H\nu_e\bar{\nu}_e}\text{BR}_{H\to Z\gamma} $  & 30.0\% \\
\hline
 & & $\sigma_{H\nu_e\bar{\nu}_e}\text{BR}_{H\to ZZ} $ & 5.6\% &  $\sigma_{H\nu_e\bar{\nu}_e}\text{BR}_{H\to ZZ} $ & 3.9\% \\
\hline
\end{tabular}
\end{center}
}
\caption{\it Summary of projected statistical precisions for Higgs measurements from Ref.~\cite{CLICH} that we use in our fit.
Observables sourced elsewhere are discussed in the text.}
\label{tab:higgsmeasurements}
\end{table}

For each of these three energy stages we use projected constraints on the Higgs measurements from Ref.~\cite{CLICH}, 
as summarised in Table~\ref{tab:higgsmeasurements}. The Higgsstrahlung process dominates most channels at 
350 GeV, whereas vector-boson fusion (VBF) provides more statistics at 1.4 and 3 TeV, providing
opportunities to identify and measure a wide range of Higgs decays. 
For the missing $H \to \gamma\gamma$, $ZZ$ and $Z\gamma$ projections at 350 GeV 
we have treated them as in Ref.~\cite{ILCwhitepaper}, assuming similar errors as for the ILC at 250 GeV. 
This is a good assumption if the errors in the branching ratios scale with the available number of Higgs bosons, 
which are similar for the ILC with 250 $\text{fb}^{-1}$ at 250 GeV (assuming 80\% polarization of the $e^-$
beam and 30\% polarization of the $e^+$ beam) to CLIC with 500 $\text{fb}^{-1}$ at 350 GeV
(with no polarization assumed).

Projections for some additional observables are needed for the analysis presented in this paper. 
Although VBF Higgs production dominates at 1.4 and 3 TeV, one of the main points of this work is to 
highlight the effect of including $HZ$ associated production at high energies,
and its importance for improving the sensitivity to certain dimension-6 operators. Also,
the $e^+ e^- \to W^+W^-$ process is important for constraining triple-gauge couplings that are important
in global fits, as noted in Table~\ref{tab:dim6}. For these reasons, additional estimates have been made for 
the $HZ$ and $W^+ W^-$ processes at generator level, including smearing and assuming 
the expected detector resolutions, where appropriate. These studies are summarised in the following. 
Confirmation of these results with full detector simulations and the study of potential systematic 
uncertainties are left for future analysis by the CLICdp Collaboration.

The cross sections and event samples were obtained using 
the {\tt WHIZARD~1.95} Monte Carlo program~\cite{arXiv:0708.4233, arXiv:hep-ph/0102195}. The effects of 
initial-state radiation (ISR) and beamstrahlung were included in the generation. The expected precisions are 
normalised to an integrated luminosity of 500~fb$^{-1}$/1.5~ab$^{-1}$/2~ab$^{-1}$ at $\sqrt{s}$=350~GeV/1.4~TeV/3~TeV.

\subsection{Higgsstrahlung at High Energy}
\label{sec:Higgsstrahlung}

The process $e^+e^- \to HZ, Z \to q\bar{q}, H \to b\bar{b}$ was chosen to evaluate the expected uncertainty in 
the Higgsstrahlung cross section at 1.4 and 3 TeV, as it provides the largest event sample. All other processes resulting in four 
final-state quarks were considered as background. The four vectors of all final-state quarks were smeared assuming 
an energy resolution of: $\sigma(E)/E = 4\%$, which corresponds to the jet energy resolution of the CLIC detector models. 
The most probable $Z$ and Higgs candidate in each event was selected by minimising:
\begin{equation*}
\chi^2 = (m_{ij} - m_Z)^2 / \sigma_Z^2 + (m_{kl} - m_H)^2 / \sigma_H^{2},
\end{equation*}
where $m_{ij}$ and $m_{kl}$ are the invariant masses of the quark pairs used to reconstruct the $Z$ and Higgs boson 
candidates, respectively, and $\sigma_{Z,H}$ are the estimated invariant mass resolutions for hadronic decays of
the  $Z$ and Higgs bosons. 
Events with a $\chi^2$ minimum of less than 20 were considered further.

Two b-tags were required for each event. For this purpose, a b-tagging efficiency of 
$80\%$/$10\%$/$1\%$ was assumed for beauty/charm/light quarks 
in the final state, as motivated by detailed simulations of realistic CLIC detector models. In order to exclude events with a large energy 
loss due to ISR or beamstrahlung, the $HZ$ invariant mass was required to be larger than 1.3/2.8~TeV 
for a centre-of-mass energy 
of 1.4/3~TeV. After the selection described above, the following uncertainties were obtained:
\begin{align*}
\frac{\Delta[\sigma(HZ) \cdot BR(H \to b\bar{b})]}{[\sigma(HZ) \cdot BR(H \to b\bar{b})]} &= 3.3\%~\text{at}~1.4~\text{TeV}, \\
\frac{\Delta[\sigma(HZ) \cdot BR(H \to b\bar{b})]}{[\sigma(HZ) \cdot BR(H \to b\bar{b})]} &= 6.8\%~\text{at}~3~\text{TeV}.
\end{align*}

\subsection{Diboson production $e^+e^- \to W^+W^-$}
\label{sec:diboson}

The most promising final states to measure the cross section for the process $e^+e^- \to W^+W^-$ 
at CLIC are $q\bar{q}q\bar{q}$ and $q\bar{q}l^\pm\nu$, 
where $l^{\pm}$ is an electron or muon. We assume that the background processes can be suppressed
to a negligible level for a signal selection efficiency of 50\% in both cases. Only events with a $W^+W^-$ invariant 
mass above 330~GeV/1.3~TeV/2.8~TeV have been considered for $\sqrt{s}=$350~GeV/1.4~TeV/3~TeV,
in order to exclude events in which ISR or beamstrahlung has a large impact. 
The following precisions are expected when combining both final states:
\begin{align*}
\frac{\Delta\sigma(W^+W^-)}{\sigma(W^+W^-)} &= 0.1\%~\text{at}~350~\text{GeV}, \\
\frac{\Delta\sigma(W^+W^-)}{\sigma(W^+W^-)} &= 0.2\%~\text{at}~1.4~\text{TeV}, \\
\frac{\Delta\sigma(W^+W^-)}{\sigma(W^+W^-)} &= 0.3\%~\text{at}~3~\text{TeV}. \\
\end{align*}

\section{Projections of CLIC Sensitivities}

We now present the potential sensitivities to the coefficients of dimension-6 operators 
that could be provided by the projected measurements and their errors outlined in the previous section. 
There are no detailed studies of the sensitivity of CLIC to electroweak precision tests (EWPTs),
so we omit such a projection here, referring instead to our previous work on the effects of 
dimension-6 operators on a leptonic subset of observables in future EWPTs~\cite{EY}.

The main operators affecting Higgs physics and triple-gauge couplings that are of interest
for our anlysis are $\bar{c}_{W}$, $\bar{c}_{HW}$, $\bar{c}_{HB}$, $\bar{c}_{3W}$, $\bar{c}_\gamma$ and $\bar{c}_g$. 
We have calculated the linear dependences of the $HZ$ Higgsstrahlung associated production cross-sections
at 1.4~TeV and 3~TeV on these dimension-6 operator coefficients using {\tt MadGraph5}~\cite{MG5},
with the following numerical results: 
\begin{align}
\left. \frac{\Delta\sigma(HZ)}{\sigma(HZ)}\right|_\text{350 GeV} &= 16\bar{c}_{HW} + 4.7\bar{c}_{HB} + 35\bar{c}_W + 11\bar{c}_B - \bar{c}_H + 5.5\bar{c}_\gamma \, , \nonumber \\
\left. \frac{\Delta\sigma(HZ)}{\sigma(HZ)}\right|_\text{1.4 TeV} &= 440\bar{c}_{HW} + 130\bar{c}_{HB} + 470\bar{c}_W + 121\bar{c}_B - \bar{c}_H + 7.3\bar{c}_\gamma \, , \nonumber \\
\left. \frac{\Delta\sigma(HZ)}{\sigma(HZ)}\right|_\text{3 TeV} &= 2130\bar{c}_{HW} + 637\bar{c}_{HB} + 2150\bar{c}_W + 193\bar{c}_B - \bar{c}_H + 7.4\bar{c}_\gamma \, .
\label{eq:HZcrosssections}
\end{align}
Similarly, in the cases of the $e^+e^- \to W^+W^-$ production cross-sections at 350 GeV, 1.4 TeV and 3 TeV, 
we obtain the following numerical results for the linear dependences on the dimension-6 operator coefficients: 
\begin{align}
\left. \frac{\Delta\sigma(W^+W^-)}{\sigma(W^+W^-)}\right|_\text{350 GeV} &= 0.63\bar{c}_{HW} + 0.31\bar{c}_{HB}  + 4.6\bar{c}_W - 0.43\bar{c}_{3W} \, , \nonumber \\
\left. \frac{\Delta\sigma(W^+W^-)}{\sigma(W^+W^-)}\right|_\text{1.4 TeV} &= 3.8\bar{c}_{HW} + 2.2\bar{c}_{HB} + 7.9\bar{c}_W - 0.66\bar{c}_{3W} \, , \nonumber \\
\left. \frac{\Delta\sigma(W^+W^-)}{\sigma(W^+W^-)}\right|_\text{3 TeV} &= 13\bar{c}_{HW} + 7.8\bar{c}_{HB} + 17\bar{c}_W - 4.4\bar{c}_{3W} \, .
\label{eq:WWcrosssections}
\end{align}
We see that the sensitivities to most of the operator coefficients increase substantially with the
centre-of-mass energy, for both the Higgsstrahlung and $W^+ W^-$ cross-sections,
confirming the expected competitive advantage of the high energies attainable with CLIC.

\begin{figure}[h!]
\begin{center}
\includegraphics[scale=0.3]{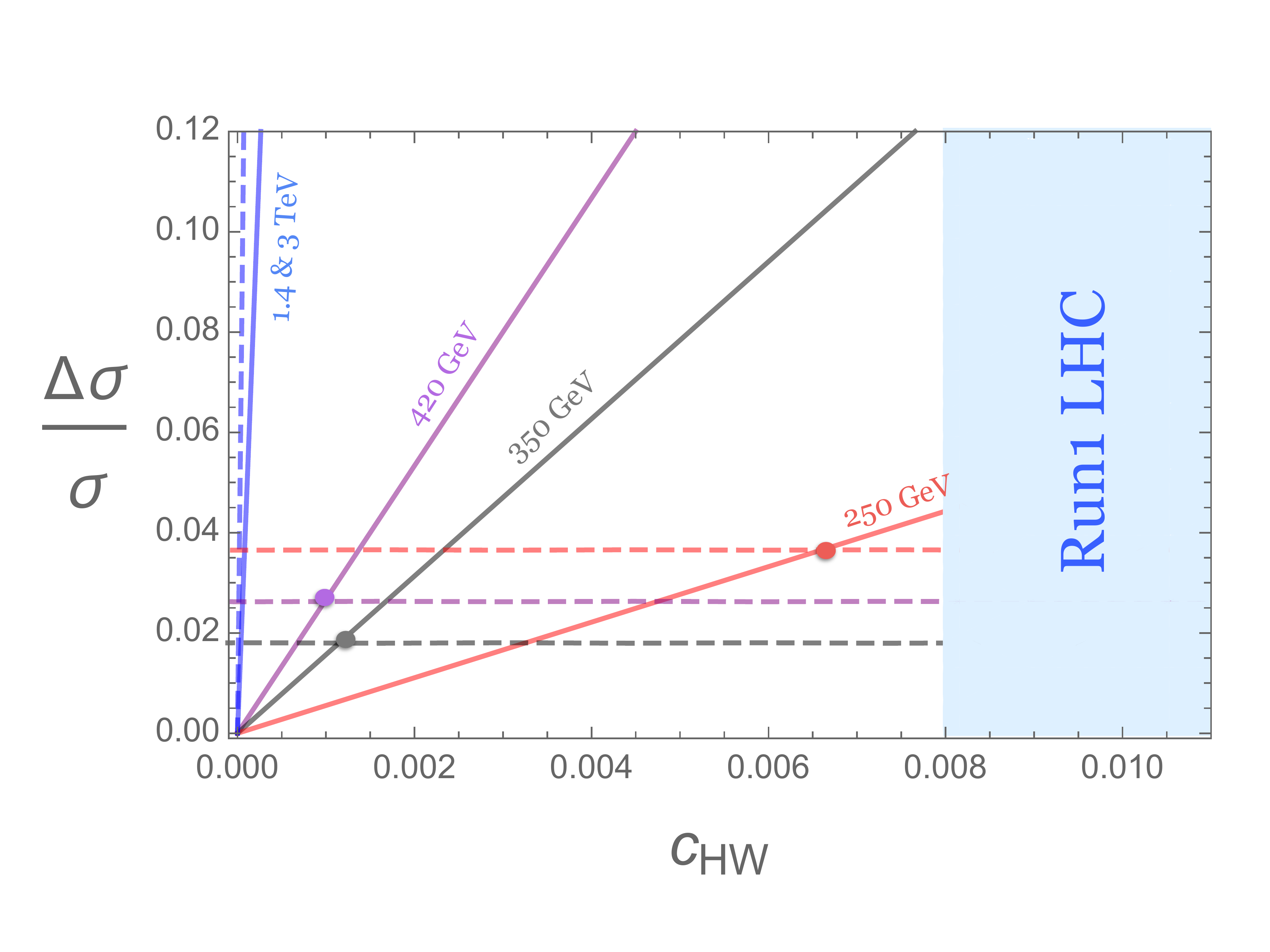}
\caption{\it   Illustration of the prospective CLIC sensitivities to the dimension-6 operator coefficient
$c_{HW}$ that could be obtained from measurements of the $HZ$ Higgsstrahlung cross-section
at different centre-of-mass energies. The diagonal solid lines represent the linear dependences
shown in Eqs.~(4.1, 4.2) and (4.3), and the horizontal dashed lines represent the measurement
errors estimated in Section~3. The range excluded by data from Run~1 of the LHC is shaded blue.}
\label{fig:cHW}
\end{center}
\end{figure}

In particular, the Higgsstrahlung cross-section has a very strong dependence on energy 
for the dimension-6 operator coefficients $\bar{c}_W$, $\bar{c}_{HW}$ and $\bar{c}_{HB}$, namely a factor
${\cal O}(100)$ between 350~GeV and 3~TeV. As an illustration, Fig.~\ref{fig:cHW} displays
the CLIC sensitivities to $\bar{c}_{HW}$ for various centre-of-mass energies, also including
estimates for 250~GeV and 420~GeV~\cite{otherCLIC} that we do not include in our analysis. 
The horizontal dashed lines denote the corresponding projected experimental errors in the measurements, 
based on the estimates in Section~3 for 1.4 and 3 TeV and the recoil cross-section measurement for the lower energies, and the dots indicate the sensitivities to $\bar{c}_{HW}$ that
could be obtained by CLIC runs at these different energies, assuming that this is the only significant
SM EFT effect. For comparison, the range of $\bar{c}_{HW}$ values excluded by the individual limit
obtained from Run 1 LHC data~\cite{ESY} is shaded in blue. We see that CLIC would improve 
marginally on this limit by running at 250~GeV, with substantial improvements possible with
higher-energy runs.

\begin{figure}[h!]
\begin{center}
\includegraphics[scale=0.5]{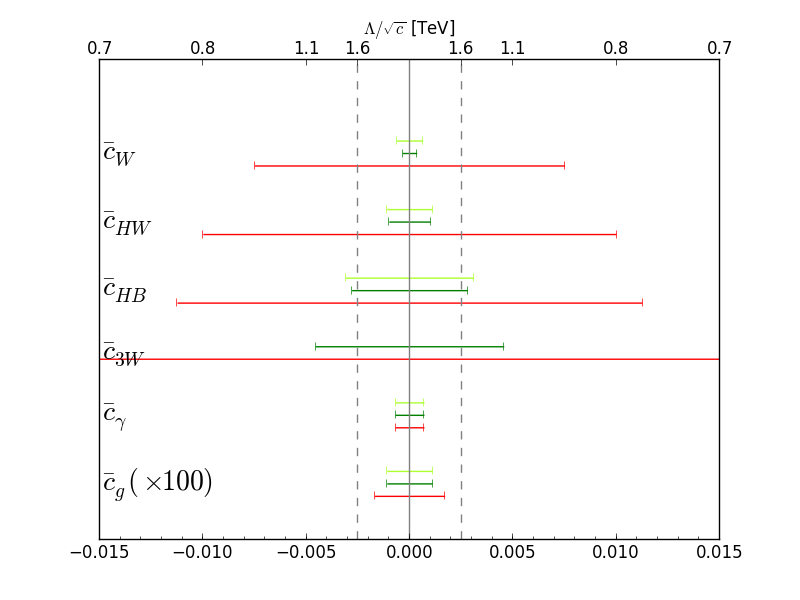}
\caption{\it  The prospective sensitivities of CLIC measurements at 350~GeV to individual
operator coefficients (green) and in a fit marginalised over all contributing operators (red).
The lighter green colour is for fits omitting the $W^+W^-$ production cross-section.}
\label{fig:CLIC350}
\end{center}
\end{figure}

\begin{figure}[h!]
\begin{center}
\includegraphics[scale=0.468]{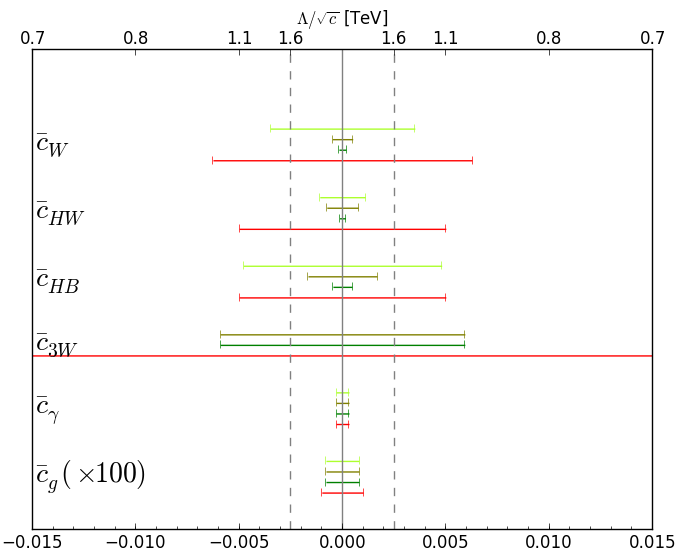}
\caption{\it  The prospective sensitivities of CLIC measurements at 1.4~TeV to individual
operator coefficients (green) and in a fit marginalised over all contributing operators (red).
The lightest green colour is for individual fits omitting both the $HZ$ and $W^+W^-$ production cross-sections,
those in olive green exclude $HZ$ but include $W^+ W^-$, and the darkest green colour is for
individual fits including both $HZ$ and $W^+W^-$.}
\label{fig:CLIC1400}
\end{center}
\end{figure}

\begin{figure}[h!]
\begin{center}
\includegraphics[scale=0.468]{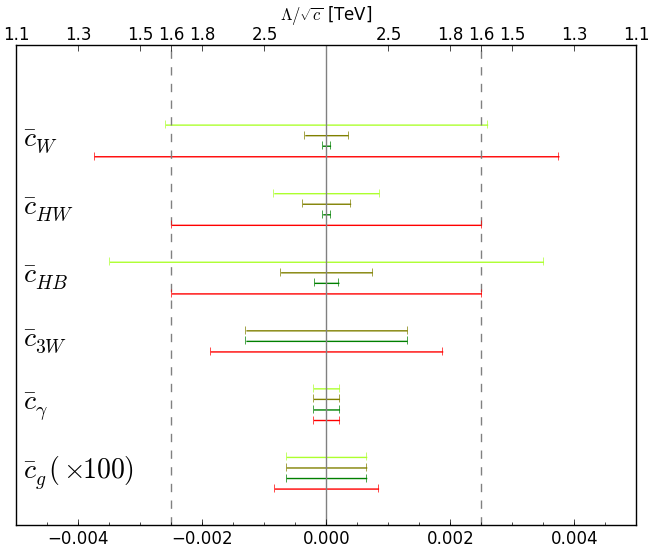}
\caption{\it  As for Fig.~\protect\ref{fig:CLIC1400}, but for CLIC measurements at 3~TeV.}
\label{fig:CLIC3000}
\end{center}
\end{figure}

Including all the various Higgs channels of Table~\ref{tab:higgsmeasurements}, we have made global fits to estimate the sensitivities
at the various CLIC centre-of-mass energies, using the dependences of the Higgs 
branching ratios on the dimension-6 operator coefficients provided by {\tt eHDECAY}~\cite{eHDECAY}
in combination with the $HZ$ production cross-section and the constraints on triple-gauge couplings from 
the $W^+W^-$ production cross-section. In making our $\chi^2$ fits, we assume Gaussian statistical errors and 
neglect theoretical uncertainties. The resulting 95\% CL limits are plotted in Figs.~\ref{fig:CLIC350},
\ref{fig:CLIC1400}, and \ref{fig:CLIC3000} for 350 GeV, 1.4 TeV and 3 TeV respectively.

The results from our 350~GeV fit shown in Fig.~\ref{fig:CLIC350} include the individual limits obtained
considering just one operator at a time colour-coded in green and the marginalised limits obtained
including all operators in red. The lighter green bounds exclude the constraints coming from $W^+W^-$ production. 
Comparing with the darker green bars that include $W^+W^-$ production, we see that this process can help
provide significantly stronger individual limits than can be obtained from Higgs physics alone. Moreover,
we note that the $\bar{c}_{3W}$ coefficient only affects triple-gauge couplings, so the inclusion of $W^+ W^-$
production is crucial in closing directions of limited sensitivity in marginalised fits. All Higgs observables 
and the $W^+ W^-$ cross section are included in the marginalised fit where all coefficients are 
allowed to vary simultaneously (including $\bar{c}_f, \bar{c}_H$, which we omit from these plots as the
constraints on them are an order of magnitude worse). 

The upper horizontal axis in Fig.~\ref{fig:CLIC350} translates the limits on the barred coefficients to the 
corresponding scale of new physics $\Lambda$ at which new physics with unit coupling
would generate a coefficient of that size when being integrated out. We note that the actual scale 
would depend on the information encapsulated in the un-barred coefficients $c_i$, 
such as the coupling strength of the new physics and whether it is loop-induced or not. With this proviso, 
we see that the $\mathcal{O}(10^{-3})$ bounds in the 350~GeV fit translate to $\sim 1$ TeV sensitivities
for the individual fits and $\sim 800$ GeV for the marginalised fit, with the notable exceptions of 
$\bar{c}_\gamma$ and $\bar{c}_g$, which are many times more precise. We note in particular that
$\bar{c}_g$ is multiplied by 100 in Fig.~\ref{fig:CLIC350}, which is equivalent to $\Lambda$ being 
multiplied by a factor of 10. As these two coefficients characterise loop-induced processes in the Standard Model, 
they are the most precise Higgs observables, and can place strong indirect limits on weakly-coupled 
loop-induced new physics such as stops in the MSSM.

The 1.4~TeV fits shown in Fig.~\ref{fig:CLIC1400} take into account, in addition to the Higgs channels of Table~\ref{tab:higgsmeasurements} 
and the $W^+ W^-$ cross-section of Section~\ref{sec:diboson}, also the $HZ$ Higgsstrahlung
production cross-section of Section~\ref{sec:Higgsstrahlung}. The individual limits shown in lightest green exclude 
both the $HZ$ Higgsstrahlung and $W^+ W^-$ cross-section constraints, while those shown in
olive green exclude the former but include the latter. The darkest green limits include both, 
as do the marginalised limits in red that now reach the $\sim 1$ TeV sensitivity for $\bar{c}_W, \bar{c}_{HW}$
and $\bar{c}_{HB}$. This is largely driven by the inclusion of the Higgsstrahlung observable,
which sets individual limits in the multi-TeV range for these operators. We emphasise, however, that although
the individual limits are beyond the centre-of-mass energy of 1.4~TeV, the marginalised limits allowing
all relevant operators to vary simultaneously are generally $< 1.4$~TeV. A realistic model may have
a scale in between, and it should be checked that one is still within the regime of validity 
of the SM EFT~\cite{aTGCinEFT, EFTvalidity}. 

Similar conclusions hold for the 3~TeV fits whose results are shown in Fig.~\ref{fig:CLIC1400}, 
where the increases in sensitivity are even more marked, with individual limits on
some operators reaching the 10~TeV level. This level of sensitivity becomes comparable
to that at which future electroweak precision tests  may constrain the orthogonal dimension-6 
operator combination $\bar{c}_W + \bar{c}_B$. A complete fit will then have to be made that includes
its effect simultaneously, including the RGE running neglected here, which mixes the coefficients at 
different energy scales~\cite{RGE}. 

\begin{figure}[h!]
\begin{center}
\includegraphics[scale=0.281]{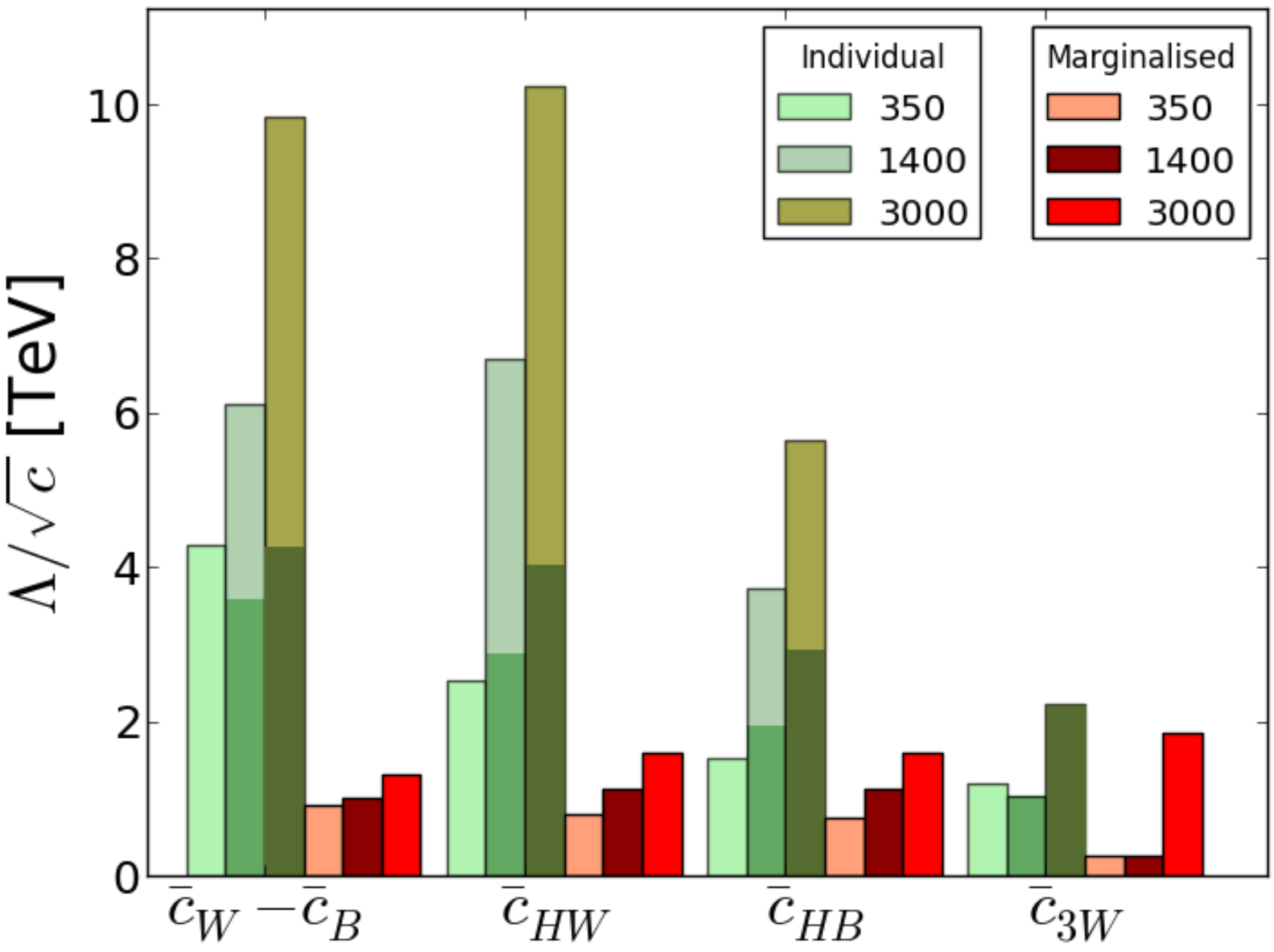}
\includegraphics[scale=0.435]{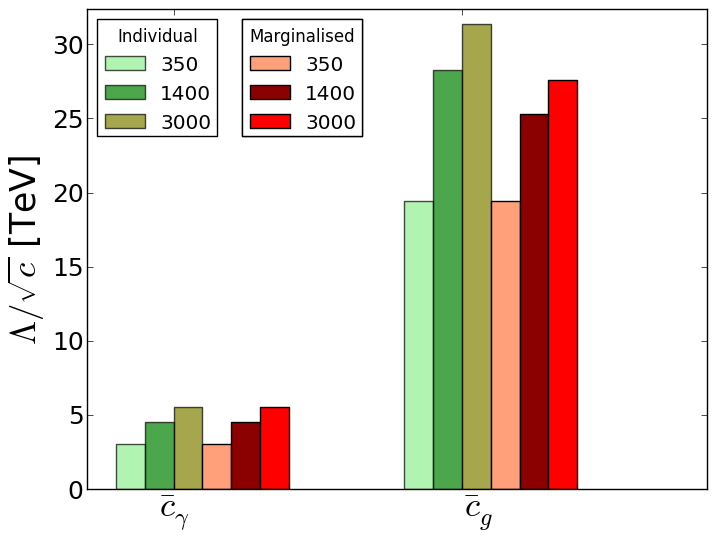}
\caption{\it  The estimated sensitivities of CLIC measurements at 350~GeV, 1.4~TeV and 3.0~TeV to the scales
of various (combinations of) dimension-6 operator coefficients: $\bar{c}_W - \bar{c}_B, \bar{c}_{HW}, \bar{c}_{HB}$
and $\bar{c}_{3W}$ (left panel) and $\bar{c}_{\gamma}$ and $\bar{c}_{g}$ (right panel). The results of
individual (marginalised) fits are shown as green (red) bars. The lighter (darker) green bars in the left panel
include (omit) the prospective $HZ$ Higgsstrahlung constraint.}
\label{fig:summary}
\end{center}
\end{figure}

The results for all three energies are summarised and compared side-by-side in the Fig.~\ref{fig:summary}, 
where the different shades of green of the individual limit bars denote the effect of including (or not)
the $HZ$ Higgsstrahlung constraint. As seen in Eqns.~\ref{eq:HZcrosssections}, the increases with energy
of the sensitivities of the operator coefficients $\bar{c}_W, \bar{c}_B, \bar{c}_{HW}$ and $\bar{c}_{HB}$
are the most rapid for this observable, and this is reflected in the increased heights of the lighter-coloured
green bars seen in the left panel of Fig.~\ref{fig:summary}. We conclude that measuring the associated
production cross-section for $e^+ e^- \to Z H$ increases significantly the CLIC sensitivities to these
operator coefficients. On the other hand, the sensitivity to $\bar{c}_{3W}$
is due exclusively to the $W^+ W^-$ cross-section observable, as seen in Eqns.~\ref{eq:WWcrosssections}. 
Finally, we note that the sensitivities to $\bar{c}_{\gamma}$ and $\bar{c}_{g}$ shown in the right panel of 
Fig.~\ref{fig:summary} increase relatively slowly with energy, as already seen in Eqns.~\ref{eq:HZcrosssections}.

Comparing with the analysis of ILC and FCC-ee sensitivities to individual SM EFT dimension-6
coefficients shown in Fig.~9 of~\cite{EY}, we see that CLIC data at 3.0~TeV would be significantly more
sensitive to $\bar{c}_{HW}$ and somewhat more sensitive to $\bar{c}_{HB}$, whereas
the CLIC sensitivities at 1.4~TeV would be comparable to those attainable at ILC/FCC-ee.
On the other hand, the CLIC sensitivity to $\bar{c}_{3W}$ is weaker than ILC/FCC-ee. 
In the cases of $\bar{c}_{\gamma}$ and $\bar{c}_{g}$, we see that CLIC could impose
stronger constraints than the ILC, but weaker than FCC-ee.

%%%%%%%%%%%%%
\section{The Reach of CLIC for Specific UV Scenarios}
%%%%%%%%%%%%%%

In order to contextualize the precision achieved by CLIC for the coefficients of the SM EFT operators, 
one may consider some specific models that could source the EFT coefficients.
Accordingly we discuss in this Section two archetypical examples of UV completions of the EFT Lagrangian, 
specifically theories with more scalar particles beyond the Higgs boson, namely stops in the MSSM and a
dilaton/radion model.  

The phenomenology of these models has been widely studied, as well as their potential 
LHC signatures: see, e.g.,~\cite{DEQY} for a discussion of indirect constraints on stops
with an explicit comparison of SM EFT and exact Feynman diagram calculations.
In this connection, we note that direct searches for stops and other new scalars are necessarily
model-dependent in nature, so that their reach is limited to specific assumptions and areas of
parameter space, whereas indirect constraints are insensitive to assumptions about their
production and decay modes. Indirect probes for new physics do not rely on the same set of 
assumptions as direct searches, and hence are a source of complementary information 
as well as a different way of finding new physics.

Virtual effects of stops in Higgs production via gluon fusion can be parametrized as an overall re-scaling of the rate, namely
\bear
\frac{\sigma ( g g \to h  )}{\sigma ( g g \to h  )_{SM}} = \kappa_g^2 \ .
\eear
Loops of stops would induce modifications in the production rate of the Higgs as follows~\cite{NSUSY}:
\bear
\kappa_g = 1 + C_g (\alpha_s) \frac{F^{SUSY}_g}{F_g^{SM}}
\eear
where the SM loop contribution is proportional to $ F_g^{SM} \simeq  -2/1.41$, and $C_g(\alpha_s) = 1+ \frac{25 \, \alpha_s}{6 \, \pi}$. 
The function $F_g$ encodes the effect of stops in loops, and is a function of the stop masses ($m_{\tilde t_{1,2}}$) and 
the mixing angle $\theta_t$ between the two chirality eigenstates:
\bear
F_g = -\frac{1}{3} \, \left[\frac{m_t^2}{m_{\tilde{t}_1}^2} + \frac{m_t^2}{m_{\tilde{t}_2}^2}  - \frac{1}{4} \, \sin^2 (2 \, \theta_t) \, \frac{\Delta m^4}{m_{\tilde{t}_1}^2 \, m_{\tilde{t}_2}^2}\right].
\eear
The best current bounds on $\kappa_g$ come from combined the Run~1 analysis of Higgs properties 
by ATLAS and CMS~\cite{Khachatryan:2016vau}, which reached a precision of about 20\%, i.e.,
$\Delta \kappa_g = |\kappa_g -1| \lesssim 0.2 $ at 95\% CL. Prospects for the High-Luminosity phase of the 
LHC  (HL-LHC) are discussed in~\cite{ATLAS-HL-LHC}, and with 3000 fb$^{-1}$ an improvement by a factor of about 3 is expected, 
leading to $\Delta \kappa_g \simeq$ 6.7\%.  

\begin{figure}[t!]
\begin{center}
\hspace{-1.cm}
\includegraphics[scale=0.25]{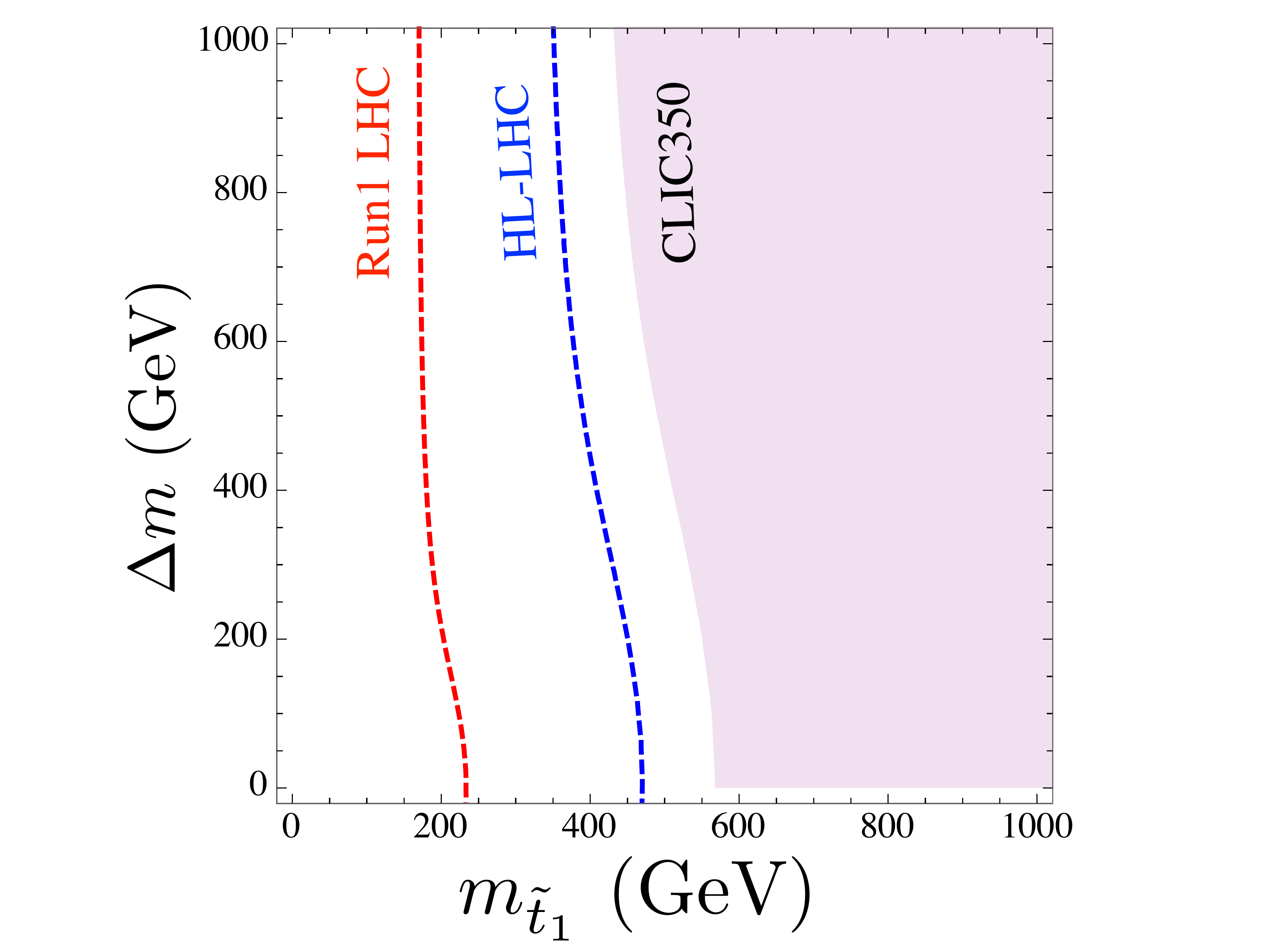}\hspace{-1.7cm}
\includegraphics[scale=0.25]{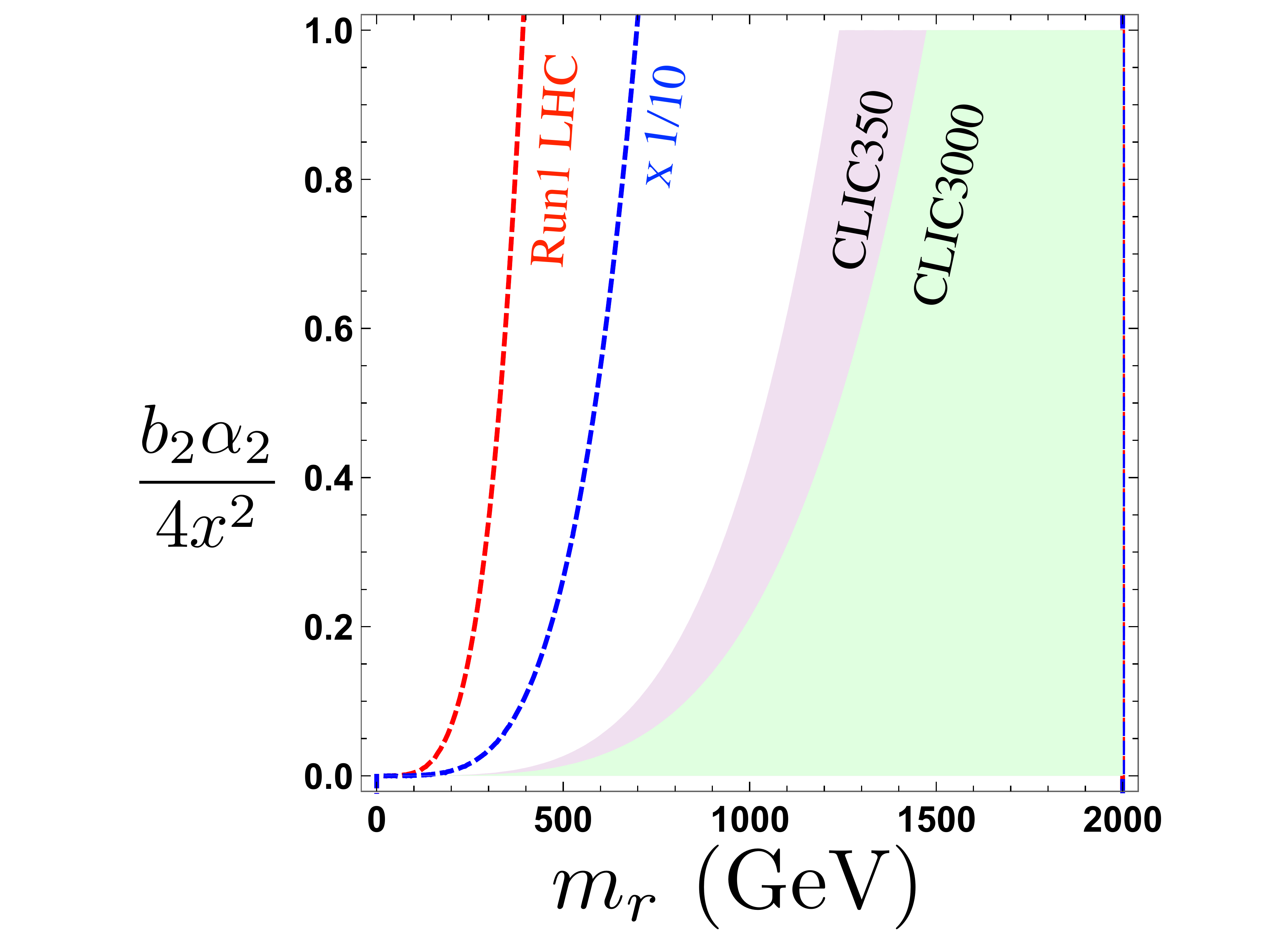}
\caption{\it  Mass reach for scenarios with new scalars. Boundaries correspond to 95\% CL exclusions. (Left panel) Current and expected mass reach for interpretations of $\bar c_g$ as limits on stop masses. (Right panel) Limits on the EFT coefficients interpreted in terms of the mass and coupling of the dilaton field, where the $\beta$-function coefficient $b_2$ and the parameter $x = f/m_r$ are introduced in the text. }
\label{fig:UVmodels}
\end{center}
\end{figure}

One can translate the limit on $\kappa_g$ in terms of the EFT coefficient $\bar c_g$ that appears in the EFT framework, as follows:
\bear
\bar c_g =  (\kappa_g-1)\frac{F_g^{SM} m_W^2}{64 \pi^2 v^2} \ .
\eear
In left panel of Fig.~\ref{fig:UVmodels} we compare the current sensitivity to MSSM stops with the
HL-LHC and CLIC prospects. We present the results in terms of the lightest stop eigenvalue $m_{\tilde t_1}$ 
and the mass separation from the heavier eigenstate $\Delta m^2 = m_{\tilde t_2}^2- m_{\tilde t_1}^2$. 
We see that the current indirect limit on a light stop of around 200 GeV will improve by a factor of two with the full HL-LHC dataset. 
This reach would be surpassed by CLIC in any energy scenario. Specifically, in the case of CLIC at 350~GeV,
we see that the sensitivity to $m_{\tilde t_1}$ is ${\cal O}(500)$~GeV, well beyond the direct production
limit of 175~GeV, so the SM EFT is quite accurate in this case~\cite{DEQY}. On the other hand, whereas the sensitivity
with CLIC at 1.4~TeV is increased to $m_{\tilde t_1} \sim 600$~GeV, this is below the threshold at which the stop is integrated out and no better than the direct
search sensitivity, while similarly the increased indirect sensitivity with CLIC at 3~TeV, $m_{\tilde t_1} \sim 700$~GeV
is significantly less than the direct kinematic reach~\footnote{We note that our results are presented for a 
specific choice of the mixing angle $\theta_t=0$, but the mass limits are roughly independent of the 
angle once other constraints such as $b\to s \gamma$ and $m_W$ are taken into account~\cite{NSUSY}.}. This is to be expected as the stops are mainly constrained by the $\bar{c}_g$ operator coefficient that does not benefit from an energy growth in the cross-section at 1.4 and 3 TeV. 

Another set of particularly interesting models is that with extended Higgs sectors, where massive scalar states could 
naturally evade discovery at the HL-LHC but lead to deviations in Higgs couplings that could be observed at CLIC. 
We focus here on the simplest such extension, namely a dilaton, or its dual, a radion scalar particle $r$.
These new particles are dual to each other, and couple to the SM via the stress tensor $T$. 

The radion/dilaton mass  is linked to the mechanism of stabilization of the extra-dimension or the explicit breaking of dilatation symmetry. 
The result is very scenario-dependent, and it could be very light as well as around the scale of compactification (or spontaneous breaking) $f$, see e.g.~\cite{GW,GW-pheno}. 
Integrating $r$ out produces the effective Lagrangian one obtains the following effective Lagrangian~\cite{ExtH}
\bear
 {\cal L}_{eff} = - \frac{1}{f^2} \frac{1}{m_r^2} T^2
 \eear  
where $T$ is the trace of the stress-energy tensor.
The relevant terms in the stress tensor  for the Higgs and gauge bosons are
  \bear
  T \subset -2 \, |D_\mu \Phi|^2 + 4 \, V(\Phi^\dagger\Phi) - \frac{b_i \alpha_i}{8 \pi}\,F^i_{\mu\nu} F^{i\mu\nu} \ ,
  \eear
where $V=-m_h^2 |\Phi|^2+\lambda |\Phi|^4$ and the $b_i$ are the $\beta$-function coefficients,
which lead to anomalous violations of scale invariance. The values of the $b_i$ depend on the degree of 
compositeness of fermions in the SM and possible new physics contributions.  For example, in conformal field theories (CFTs) 
one typically obtains, $b^{CFT}_i=8 \pi^2/(g_i^2 \log(\mu_{IR}/\Lambda_{UV}))$~\cite{GW-pheno}, in which case 
\bear
b_2 \alpha_2 \simeq 2 \pi \log(\mu_{IR}/\Lambda_{UV}) \ .
\eear
One can easily read off the following coefficients of dimension-6 operators:
\bear
 \bar c_{HW} = - \bar c_W  & =&  - \frac{b_2 \alpha_2}{4}  \, \frac{m_h^2 v^2}{f^2 m_r^2} \, , \nonumber \\
\bar c_{HB} = - \bar c_B  & = & - \frac{b_1 \alpha_1}{4}  \, \frac{m_h^2 v^2}{f^2 m_r^2} \, .
\eear
One could also consider a more general situation with a non-universal dilaton/radion coupling. 
This  would lead to a prefactor in the coefficients of the effective operators, dependent on the degree of 
overlap of the wavefunctions in the bulk (radion) or participation on the composite dynamics of the species (dilaton),
but we do not enter into details here. 

The dominant dilaton/radion production mechanism at CLIC would be in association with a $Z$ boson, 
$e^+ e^- \to Z \to Z r$. In the case of CLIC at 350~GeV, the production cross-section scales as $\sim$1 pb (1 TeV$/f)^2$ 
for $m_r \lesssim $ 200 GeV, and at 3.0~TeV CLIC would produce 1-TeV radions with a cross-section 0.5 pb (1 TeV$/f)^2$.  
In models where the composite dynamics is related to the origin of electroweak symmetry breaking, 
one finds that the radion/dilaton decays predominantly to massive particles whenever kinematically allowed, 
with ratios $2:1:1$ to the final states $W^+ W^-$, $ZZ$ and $hh$ respectively~\cite{maxime}. 
Therefore the signatures at the LHC are dominated by diboson decays, which are most
distinctive in the high-mass region (above a TeV). 
 
We show in the right panel of Fig.~\ref{fig:UVmodels} indirect limits on the radion parameter space via the effect on the 
SM EFT with the constraints $\bar c_{HW} = -\bar c_W$ and $\bar c_W + \bar c_B \simeq 0$. 
We also compare with the limits from the Run-I fit~\cite{ESY} and an optimistic ballpark estimate of
an improvement of a factor ten for $\bar c_{HW}$ in the high-luminosity LHC runs.  
We see that CLIC can extend the indirect reach for the radion mass into the TeV scale, greatly
exceeding the direct reach of the LHC. However, we note that for CLIC at 3~TeV the exclusion contour extends
into a region where the EFT expansion breaks down, though there could still be indirect probes of the 
radion via the behaviour of the Higgs.

As far as direct searches are concerned, we note that a non-universal dilaton/radion could decay predominantly to an invisible final state
or one that is particularly difficult to detect. At 3 TeV, the cross section for $e^+ e^-$
to neutrinos and a $Z$ boson is 2~pb, much larger than the cross section expected in the dilaton/radion model considered here. 
So, if the radion/dilaton decays into invisible (undetectable) final states, further kinematic selections would be needed to 
distinguish new phenomena from the neutrino background. This example illustrates the complementarity between 
indirect and direct probes, as they are based on different assumptions about the couplings to SM particles and dominant branching ratios. 

The gain in sensitivity for CLIC at 1.4 and 3 TeV comes from the energy dependence of certain operators 
that can place tight constraints on their Wilson coefficients. However, if the mass scale of the UV physics being integrated out 
lies below the kinematic threshold then the parameter space of such models is outside the regime of EFT validity. 
On the other hand, in some strongly-coupled models the UV mass scale may lie significantly
beyond this validity threshold~\cite{EFTvalidity, compositefermions}. This may occur if there is a strong coupling whose degeneracy with the UV
mass scale can maintain a fixed value for the Wilson coefficient's contribution to the cross-section. 
For example, Ref.~\cite{compositefermions} discusses a model of composite fermions whose effects in Higgsstrahlung 
can be generated by $\bar{c}_W - \bar{c}_B$ via such a strong coupling. Such a scenario would be 
beyond the direct reach of CLIC but may still be probed indirectly in the higher-energy runs
with an SM EFT analysis.

\section{Summary and Conclusions}

We have emphasized in this paper the potential importance of measurements of $e^+ e^- \to H Z$,
$H \nu {\bar \nu}$ and $W^+ W^-$ in high-energy CLIC running for indirectly probing possible new physics
beyond the SM by constraining the coefficients of dimension-6 operators in the SM EFT. We have
stressed, and shown numerically, that the increased relative importance of interferences between
SM and dimension-6 EFT amplitudes at high energies provides opportunities in the processes
$e^+ e^- \to H Z$ and $W^+ W^-$, in particular.

These processes have not yet been simulated in detail by the CLICdp Collaboration~\cite{CLIC, CLICH},
so we have presented estimates of the prospective precision with which their cross-sections could
be measured in CLIC running at 1.4 and 3 TeV. We have then incorporated these estimates in fits
of the relevant SM EFT coefficients, showing that such measurements are orders of magnitude
more sensitive than low-energy measurements. In some cases, the sensitivity to new physics in
individual dimension-6 operator coefficients may reach the level of 10 TeV, greater than that
attainable with the ILC of FCC-ee~\cite{EY}. 

We matched the coefficients on to specific UV models and analysed the implications for the 
indirect sensitivity to new particles in these scenarios. We find that CLIC at 350 GeV may 
outperform the indirect constraints from LHC and HL-LHC on stops and radions/dilatons. 
The 1.4 and 3 TeV runs, on the other hand, are uniquely sensitive to operators whose contributions
to the Higgsstrahlung process grows with energy. They may be used to constrain specific 
strongly-coupled models of composite fermions that are inaccessible to direct searches. 

Our results motivate more detailed studies including additional benchmark analyses based
on full CLIC detector simulations at high energies,
with the aim of verifying and refining our estimates of the accuracies with which the 
cross-sections for $e^+ e^- \to H Z$ and $W^+ W^-$ could be measured at CLIC. It has
long been clear that the higher centre-of-mass energies attainable with CLIC offer significant
advantages in direct searches for heavy particles that appear in scenarios for physics beyond
the SM. The results of our paper indicate that the same should be true for indirect searches
for such new physics.

\section*{Acknowledgements}

The research of JE has been supported partly by the Science Technology and Facilities Council (STFC) under grant ST/L000326/1,
and the research of VS has been supported partly by the STFC under grant ST/J000477/1. TY is supported by a Research Fellowship from 
Gonville and Caius College, Cambridge.

\newpage

 \providecommand{\href}[2]{#2}\begingroup\raggedright

\end{document}